\documentclass[twocolumn,preprintnumbers,amstex]{revtex4}
\usepackage{amsmath}
\usepackage{graphicx, color}
\usepackage{dcolumn}
\usepackage{bm}
\usepackage[english]{babel}
\usepackage{t1enc}
\usepackage{subfigure}

\begin{document}

\title{Spin selective transport through Aharonov-Bohm and Aharonov-Casher triple quantum dot systems}
\author{Leandro Tosi}
\author{A.~A.~Aligia}
\email{aligia@cab.cnea.gov.ar}
\affiliation{Centro At\'{o}mico Bariloche and Instituto Balseiro, Comisi\'{o}n Nacional
de Energ\'{\i}a At\'{o}mica, 8400 Bariloche, Argentina}
\date{\today}

\begin{abstract}
We calculate the conductance through a system of three quantum dots under two
different sets of conditions that lead to spin filtering effects under an applied 
magnetic field. In one of them, a spin is localized in one quantum dot, 
as proposed by Delgado {\it et al.} [Phys. Rev. Lett. \textbf{101}, 226810 (2008)].
In the other one, all dots are equivalent by symmetry and the system is subject 
to a Rashba spin-orbit coupling. 
We solve the problem using a simple effective
Hamiltonian for the low-energy subspace,  improving the accuracy of previous results.
We obtain that correlation effects
related to the Kondo physics play a minor role for parameters estimated previously
and high enough magnetic field.
Both systems lead to a magnetic field tunable ``spin valve''.
\end{abstract}

\maketitle

\section{Introduction}

Quantum dots (QD's) have attracted much attention recently because of its
possible numerous applications and the impressive experimental control
available in manipulating microscopic parameters. It has been established
that they are promising candidates in quantum information processing, with
spin coherence times of several microseconds \cite{pett,grei} and fast
optical initialization and control \cite{ata,grei2,bere}. QD's are also of
interest in the field of spintronics (electronics based on spin) 
\cite{mac,kork,rei}. Systems with one QD \cite{gold1,cro,gold2,wiel} or one
magnetic impurity on the (111) surface of noble metals \cite{man}, were used
to test single impurity models with strong correlations, in which the Kondo
physics was clearly displayed, confirming predictions based on the Anderson
model \cite{glaz,ng,chz,rev}. More recently, the scaling laws in transport
under an applied bias voltage in a non-equilibrium situation \cite%
{grobis,rin,scott,sela}, quantum phase transitions \cite%
{roch,rou1,logan,roch2,rou2,mitc}, and the effect of hybridization in
optical transitions \cite{ata,dal,misael} are subjects of intense research.

In addition, systems of several impurities or QD's have also attracted much
attention. For example non-trivial results for the spectral density were
observed when three Cr atoms are placed on the (111) surface of Au \cite%
{jamn,trimer}. Systems of two \cite{craig,hol,jeong,chen,vidan,petta}, three 
\cite{waugh,gaud,gaudpss,gaudprb,grove,schro,rogge}, and more \cite{kouw}
QD's have been assembled to study the effects of interdot hopping on the
Kondo effect, and other physical properties driven by strong correlations.
Particular systems have been proposed theoretically to tune the density of
states near the Fermi energy \cite{dias,oje}, or as realizations of the
two-channel Kondo model \cite{oreg,zit}, the double exchange mechanism \cite%
{mart}, and the so called ionic Hubbard model \cite{ihm}. Transport through
arrays of a few QD's \cite{ihm,corn,zit2,ogur,nisi,3do,kuzme,dinu,ldeca,abc,abc2,del,del2,del3,mitc2}
and spin qubits in double QD's \cite{sq} have been studied theoretically.
 In particular, the conductance through a system of three QD's described
by the Hubbard model has been studied, as a function of a gate voltage that
changes the occupation in the system gradually from 2 to 4 electrons, in a
linear array and in an isosceles geommetry \cite{3do}. At low temperatures,
as a consequence of the development of the Kondo effect, the conductance
reaches the ideal value $2e^{2}/h$ when the occupation of the system is near
3. The amplitude of the plateau of ideal conductance (the range of gate
voltage with odd occupancy) is however smaller than the case of only one QD 
\cite{3do}. For an equilateral shape and an odd number of electrons in the
system, there are additional degeneracies that lead to interesting physics 
\cite{kuzme,mitc2}. More recently, the conductance through similar systems
was studied for a fewer number of electrons, with methods which are not able
to capture the Kondo physicas, as described below \cite{del,del2,del3}. 

Another subject of interest are effects of interference in quantum paths and
the Aharonov-Bohm effect. They have been demonstrated in mesoscopic rings
with embedded QD's  in transport \cite{wiel,hol,zaf,ji} and in optical 
\cite{bayer,teodo} measurements. Recent experiments in semiconductor
mesoscopic rings \cite{kon,ber} have shown that the conductance oscillates
not only as a function of the applied magnetic field (Aharonov-Bohm effect)
but also as a function of the applied electric field perpendicular to the
plane of the ring (Aharonov-Casher effect \cite{ac}). In this effect, the
electrons, as they move, capture a phase that depends on the spin, as a
consequence of spin-orbit coupling. The main features of the experiment can
be understood in a one-electron picture \cite{shen,mol}. However, when the
spin-orbit coupling is smaller or of the order of the Kondo energy, the
conductance should be described with a formalism that includes correlations 
\cite{abc,abc2}. A device of several QD's in an Aharonov-Bohm-Casher
interferometer can be used as a spin filter, since the conductance depends
on spin and can be made to vanish for one spin direction \cite{abc,abc2}.
Similar effects were proposed in a system of two non-interacting QD's \cite%
{chi,chi2}. This is an alternative to previous proposals, such as using one
QD under an applied magnetic field \cite{torio,lady}. Similar calculations
including spin-orbit coupling have been performed \cite{sun,heary,vernek},
but a recent study indicates that this coupling works against the desired
spin-filtering effects when only one QD is present in the ring \cite{vernek}%
. Recently, it was shown that an array of QD's which combines Dicke and Fano
effects can be used as a spin filter \cite{oje}. More recently, the effects
of alternating magnetic fields on spin blockade was calculated \cite{busl}.
There are also studies in Luttinger-liquid wires with impurities \cite%
{schm,rist}.

Delgado \textit{et al.} have studied theoretically the conductance through a
system of three QD's threaded by a magnetic flux (see Fig. 1) \cite{del,del2}. 
They show that tuning the on-site energies in a particular way through
applied gate voltages (localizing a spin on QD 2), the device can be used as
a spin filter. The authors have used a sequential tunneling approximation to
calculate the conductance. This approach cannot describe the Kondo effect.
In addition, from their work, it is not clear that the conductance for each
spin $G_{\sigma }$ at zero temperature reaches the ideal value $G_{0}$ ($%
e^{2}/h$ for the symmetric case). In fact the authors assume very small
lead-dot hopping $t_{LD}$ and their result for $G_{\sigma }$ is proportional
to $|t_{LD}|^{2}$. This cannot be valid at very small temperature, for which
(as we show in Section IV), the maximum of $G_{\sigma }$ is near $G_{0}$
independently of $t_{LD}$.

In the first part of this paper, we study the same system as Delgado 
\textit{et al.}, mapping the relevant eigenstates of the system either to an
effective Anderson model or to an effective non-interacting model depending
on the parameters. The conductance of the resulting model can be evaluated
rigorously using known expressions \cite{rin,meir}. The conductance as a
function of the gate voltage or the applied flux displays spikes reaching a
value near $e^{2}/h$ when the energy necessary to transfer an electron from
the leads to the three-dot system , or conversely, vanishes. In agreement
with Ref. \cite{del}, the conductance is spin polarized. 

In the second part of the paper we include the effect of spin-orbit
coupling. We show that the spin filtering effect persists without the need
of setting different on-site energies for the three QD's, and furthermore, a
small change in the applied magnetic field reverses the spin polarization of
the current.

In Sections II and III we describe the model and approximations
respectively. Section IV contains our results. Section V contains a summary
of the results and a discussion.

\bigskip

\section{Model}

The system, represented in Fig. \ref{scheme}, consists of an array of three
QD's threaded by a magnetic flux, attached to two conducting leads, and
subject to a Rashba spin-orbit interaction. The Hamiltonian can be written
in the form 
\begin{equation}
H=H_{TQD}+H_{L}+H_{LD},  \label{hamtot}
\end{equation}%
where the first term consists of an extended Hubbard model that describes
the triple QD \cite{3do,abc,abc2,del}

\begin{eqnarray}
H_{TQD} &=&\sum_{i\sigma }\{E_{i\sigma }n_{i\sigma }-t\left[ e^{i(\Phi
_{AB}+\sigma \Phi _{AC})/N}d_{i+1\sigma }^{\dagger }d_{i\sigma }+\mathrm{H.c.%
}\right] \}  \notag \\
&&+\sum_{i} \{Un_{i\downarrow }n_{i\uparrow }+Vn_{i}n_{i+1}\},.
\label{hamtqd}
\end{eqnarray}
with $N=3$. Here $d_{i\sigma }^{\dag }$ creates an electron in the QD $i$
with spin $\sigma $, $n_{i\sigma }=d_{i\sigma }^{\dag }d_{i\sigma }$, and $%
n_{i}=n_{i\downarrow }+n_{i\uparrow }$. The phase accumulated by the hopping
terms $(\Phi _{AB}+\sigma \Phi _{AC})$, $\sigma =\pm 1$, contains an
Aharonov-Bohm phase $\Phi _{AB}=2\pi \phi /\phi _{0}$, where $\phi $ is the
magnetic flux threading the triangle and $\phi _{0}=h/e$ is the flux
quantum, as well as a spin dependent Aharonov-Casher phase \cite{abc,abc2}

\begin{equation}
\Phi _{AC}=\sqrt{\pi ^{2}+R^{2}}-\pi ,  \label{phiac}
\end{equation}%
where

\begin{equation}
R=\pi \hbar \alpha E_{z}N/(2ta),  \label{r}
\end{equation}%
with $E_{z}$ the electric field perpendicular to the plane of the triangle, $%
\alpha $ the Rashba spin-orbit coupling constant and $a$ the interdot
distance. Note that if $\alpha \neq 0$, the spin quantization axis is tilted
in the radial direction by an angle $\theta =\arctan (R/\pi )$ \cite%
{abc,abc2}. The applied magnetic field perpendicular to the plane of the
triangle originates a Zeeman term, which with the chosen quantization axis
has two terms, one that splits the on-site energies

\begin{equation}
E_{i\sigma }=E_{i}-\sigma g\mu _{B}B(\cos \theta )/2,  \label{eisigma}
\end{equation}%
and another spin-flip term that we can safely neglect for realistic
parameters, as those given in Ref. \cite{del}. $U$ and $V$ represent the
on-site and nearest-neighbor interactions respectively.

The second term in Eq. (\ref{hamtot}) describes the non-interacting leads.
Their detailed description is not important as long as they have a
featureless electronic structure near the Fermi energy. We represent them as
semi-infinite chains following previous works \cite{ihm,3do,del}

\begin{equation}
H_{L}=-t_{L}[\sum_{i=-1,\sigma }^{-\infty }\left( c_{i\sigma }^{\dag
}c_{i-1\sigma }+\mathrm{H.c.}\right) +\sum_{i=1,\sigma }^{+\infty }\left(
c_{i\sigma }^{\dag }c_{i+1\sigma }+\mathrm{H.c.}\right) ].  \label{hleeds}
\end{equation}

The remaining term in $H$ couples leads and dots 
\begin{equation}
H_{LD}=-t_{LD}(c_{-1\sigma }^{\dag }d_{1\sigma }+c_{1\sigma }^{\dag
}d_{3\sigma }+\mathrm{H.c.}).  \label{hmix}
\end{equation}

.\newline

\section{Formalism}

For week enough $t_{LD}$, as it is usually the case, an excellent
approximation consists in retaining the low-energy states of $H_{TQD}$ and
map the system into one effective dot, coupled to the leads. This approach
has been used before for similar systems \cite{ihm,3do,abc,abc2}. For
transport, the most interesting case (because it leads to a larger
conductance) is when the ground states of two neighboring configurations,
with say $n$ and $n+1$ particles are nearly degenerate. We are assuming,
without loss of generality for equilibrium, that the origin of energies is
at the Fermi level. One of the simplest mappings is realized when these
ground states correspond to a doublet and a singlet. Then, the resulting
effective model is equivalent to an impurity Anderson model with infinite
on-site repulsion \cite{ihm}. Although this model is not trivial, it can be
appropriately handled by different techniques \cite{hews,none} and the
conductance exhibits the characteristic features of the Kondo effect when
the doublet is well below the singlet \cite{none}. However, when quantum
interference effects as those present in Aharonov-Bohm rings are important,
more low-energy states should be included \cite{inter} and the effective
Hamiltonian is more involved \cite{abc,abc2}.

\begin{figure}[h]
\includegraphics[width=8cm]{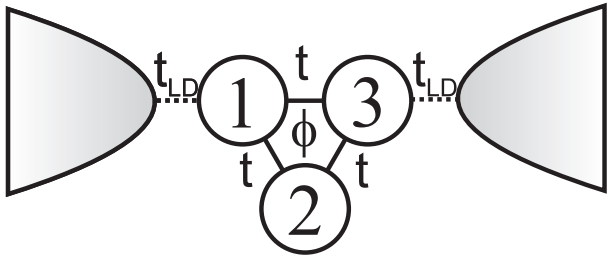}
\caption{Scheme of the system of three QD's coupled to two conducting leads.}
\label{scheme}
\end{figure}

In the present paper we assume that the gate voltage $V_{g}$ which shifts
rigidly the on-site energies $E_{i\sigma }$ by the amount $-eV_{g}$ is such
that the configurations with one and two particles are nearly degenerate.
For realistic parameters \cite{del}, we obtain that the original model can
be mapped accurately into an Anderson model. Moreover, when either the
magnetic field $B$ or the spin-orbit coupling $\alpha $ is large enough, one
can retain only \emph{one} state of both configurations and the effective
model reduces to a \emph{spinless} one, which can be solved trivially. Here
we describe in more detail this mapping, assuming that the ground state for $%
n=2 $ is connected adiabatically with a singlet for $\alpha =0$. This is
always true for not too high magnetic field \cite{note}. For simplicity, we
also assume that the system has reflection symmetry (see Fig. \ref{scheme}).
The generalization to other cases is straightforward and follows similar
procedures as in previous papers \cite{ihm,3do,abc,abc2}.

The ground state for two electrons $|2g\rangle$ is mapped into the vacuum
state $|0\rangle $ of a fictitious effective QD. Similarly, the ground state
for one electron $|1g\rangle$ is mapped into $\tilde{d}^{\dagger }|0\rangle $%
. For simplicity, in the following derivation we assume that $|1g\rangle$
has spin up (this is the case for $B>0$ and $\alpha=0$). Otherwise spin up
and down should be interchanged below. Introducing hole operators $%
h_{j}=e^{i\varphi _{j}}c_{j\downarrow }^{\dag }$ for the leads (with phases $%
\varphi _{j}$ chosen in such a way that $\tilde{t}_{LD}$ below is real and
positive), the effective Hamiltonian takes the form

\begin{eqnarray}
H_{eff} &=&t_{L}[\sum_{i=-1}^{-\infty }\left( h_{i}^{\dag }h_{i-1}+\mathrm{%
H.c.}\right) +\sum_{i=1}^{+\infty }\left( h_{i}^{\dag }h_{i+1}+\mathrm{H.c.}%
\right) ]  \notag \\
&&+E_{d}\tilde{d}^{\dagger }\tilde{d}-\tilde{t}_{LD}(h_{-1}^{\dag }\tilde{d}%
+h_{1}^{\dag }\tilde{d}+\mathrm{H.c.}),  \label{heff}
\end{eqnarray}%
where

\begin{eqnarray}
E_{d} &=&E_{1g}-E_{2g},  \notag \\
\tilde{t}_{LD} &=&t_{LD}|\langle 2g|d_{1\downarrow }^{\dagger }|1g\rangle |,
\label{map}
\end{eqnarray}%
and $E_{ng}$ is the energy of $|ng\rangle $.

Note that the electrons that can hop to the three-dot system, rendering $%
|1g\rangle$ into $|2g\rangle$, have opposite spin as $|1g\rangle$. Thus the
conduction for spin up is zero and the current is polarized down within this
approach (due to the truncation of the Hilbert space). The validity of the
approach and how this argument is modified when more states are included in
the low-energy manifold, is discussed in the next section.

The spectral density at the effective dot is given by

\begin{equation}
\rho _{d}(\omega )=\frac{i}{2\pi }\left( \langle \langle \tilde{d};\tilde{d}%
^{\dagger }\rangle \rangle _{\omega +i\epsilon }-\langle \langle \tilde{d};%
\tilde{d}^{\dagger }\rangle \rangle _{\omega -i\epsilon }\right) ,
\label{green}
\end{equation}%
and the Green's function $\langle \langle \tilde{d};\tilde{d}^{\dagger
}\rangle \rangle $ can be obtained from the equations of motion. 
Specifically, for any fermion operator $f$ one has

\begin{equation}
\omega \langle \langle f;\tilde{d}^{\dagger }\rangle \rangle =\{f,\tilde{d}%
^{\dagger }\}+\langle \langle \left[ f,H_{eff}\right] ;\tilde{d}^{\dagger
}\rangle \rangle .  \label{mov}
\end{equation}%
Starting from the case $f=\tilde{d}$, new Green's functions are generated by
the commutator (with $\tilde{d}$ replaced by $h_{1}$ and $h_{-1}$). Using
again Eq. (\ref{mov}) with $f=h_{1}$, $\langle \langle h_{2};\tilde{d}%
^{\dagger }\rangle \rangle $ appears and so on. This chain of equations can
be solved using the ansatz $\langle \langle h_{i+1};\tilde{d}^{\dagger
}\rangle \rangle =\gamma \langle \langle h_{i};\tilde{d}^{\dagger }\rangle
\rangle $, with $i\geq 1$. From Eq. (\ref{mov}) one obtains a quadratic
equation for $\gamma $ and the physical solution solution for it for
frequencies inside the band of the lead is 
\begin{equation}
\gamma =-\frac{\omega }{2t_{L}}\pm i\sqrt{1-\left( \frac{\omega }{2t_{L}}%
\right) ^{2}},  \label{gamma}
\end{equation}%
where the sign is the same as that of the positive infinitesimal in $\omega
\pm i\epsilon $. A similar solution is obtained for the operators of the
other lead ($\langle \langle h_{i};\tilde{d}^{\dagger }\rangle \rangle $,
with $i<0$). After some algebra one obtains

\begin{equation}
\rho _{d}(\omega )=\frac{4\tilde{t}_{LD}^{2}\left[ 4t_{L}^{2}-\omega ^{2}%
\right] ^{\frac{1}{2}}}{\pi \left\{ \left[ (\omega -E_{d})\omega -4\tilde{t}%
_{LD}^{2}\right] ^{2}+(\omega -E_{d})^{2}\left[ 4t_{L}^{2}-\omega ^{2}\right]
\right\} }.  \label{dens}
\end{equation}%
Using the same formalism as above, one obtains the spectral density at the
first site of a  semi-infinite chain that corresponds to an isolated lead:

\begin{equation}
\rho _{1}(\omega )=\frac{1}{\pi t_{L}}\left[ 1-\left( \frac{\omega }{2t_{L}}%
\right) ^{2}\right] ^{\frac{1}{2}}.  \label{chain}
\end{equation}

The conductance can be evaluated using Eqs. (\ref{dens}), (\ref{chain}) and
known expressions for nonequilibrium dynamics \cite{meir,none}. In the
present paper we restrict ourselves to the linear response regime and the
expression for the conductance takes the form \cite{pro}

\begin{equation}
G=G_{0}\frac{2\pi \tilde{t}_{LD}^{2}}{t_{L}}\int \left( -\frac{\partial f}{%
\partial \omega }\right) \left[ 1-\left( \frac{\omega }{2t_{L}}\right) ^{2}%
\right] ^{\frac{1}{2}}\rho _{d}(\omega )d\omega ,  \label{cond}
\end{equation}%
where $G_{0}=e^{2}/h$ and $f(\omega )$ is the Fermi function. While the
actual derivation of this equation is lengthy, the result is intuitive; the
conductance is proportional to the product of density of available states at
the leads $\rho _{1}(\omega )$, the density of states at the system $\rho
_{d}(\omega )$, and the change in the occupation of these states when a
small bias voltage $V$ is applied [$f(\omega +eV/2)-f(\omega +eV/2)\simeq
eV\partial f/\partial \omega $].

Because of the neglected excited states in the derivation of \ $H_{eff}$,
the above expression is valid as long as for each configuration, the energy
difference between excited states and the ground state is larger than $kT$
and the resonant level width $\Delta =2$ $\tilde{t}_{LD}^{2}/t_{L}$. 
This means in particular that to use Eq. (\ref{cond}) any possible Kondo
effect (which requires degenerate states) has been destroyed by magnetic
field $B$. This is the case for the parameters of the model for not too
small $B$.  When $B$ is not strong enough, the states which become Kramers
degenerate with $|1g\rangle $ for $B=0$ should both be included in $H_{eff}$
and the effective model becomes equivalent to the Anderson model under an
applied magnetic field. To treat this case, we use a slave-boson mean-field
approximation (SBMFA) \cite{ihm,lady}. The conductance at zero temperature
is given by

\begin{eqnarray}
G &=&G_{\uparrow }+G_{\downarrow },  \notag \\
G_{\sigma } &=&\frac{e^{2}}{h}\sin ^{2}(\pi n_{\sigma }),.  \label{gand}
\end{eqnarray}%
where $n_{\sigma }$ is the probability of finding the lowest lying state
with one particle and spin $\sigma $ in the ground state. Within the SBMFA,
the $n_{\sigma }$ are calculated solving a self-consistent set of equations
described in detail in Ref. \cite{lady}. Since it turns out that the
Anderson and Kondo physics plays a minor role in the present paper, we do
not reproduce these equations here.

\section{Results}

As a basis for our study we have taken the parameters given by Delgado 
\textit{et al.} \cite{del,del2}, calculated using a method based on linear
combination of harmonic orbitals and configuration interaction \cite{gim}.
We set $t\simeq 0.23$ meV as the unit of energy, $U=50t$, $V=10t$, $%
t_{L}=100t$ and $g=0.44$. For $t_{LD}$ we have taken several values of the
order of a few times $t$, leading to a resonant level width $\Delta \sim
0.01t$.

We have performed two sets of calculations. In the first one, the one-site
energies $E_{i}$ are tuned in such a way that the isolated system 
(for $t_{LD}=0$) lies near a quadruple degeneracy point, as described by Delgado 
\textit{et al. } \cite{del,del2} and the spin-orbit coupling $\alpha =0$. In this
case, $E_{1}=E_{3}$, and $E_{2}-E_{1}=V-U$. In the second set, the isolated
system has $C_{3}$ symmetry, and therefore, $E_{1}=E_{2}=E_{3}$, but $\alpha
\neq 0$. Using an interdot distance $a=60$ nm \cite{del} and $t=0.23$ meV,
Eq. (\ref{r}) gives $R\simeq \hbar \alpha E_{z}/$(3 nm meV). Typical values 
\cite{usaj} of $\hbar \alpha E_{z}$ with an intrinsic electric field $E_{z}$
are 0.5 nm meV for GaAs \cite{miller} and 10 nm meV for InAs \cite%
{nitta,grun}. Using externally applied $E_{z}$, the value of $R$ can be
increased up to 5 \cite{kon,ber}. Here we take $R$ of the order of 1, for
which the mapping to an effective spinless model is valid in general \cite%
{note2}. For $R$ of the order of \ 0.1 or less, more than one state in the
one-particle subspace becomes relevant and the effective Hamiltonian is more
involved \cite{abc,abc2}.

\subsection{System with different on-site energies without spin-orbit
coupling}

Here we report the results with $\alpha =0$, $E_{1}=E_{3}$, and $%
E_{2}-E_{1}=V-U$.

\medskip

\begin{figure}[h]
\includegraphics[width=8cm]{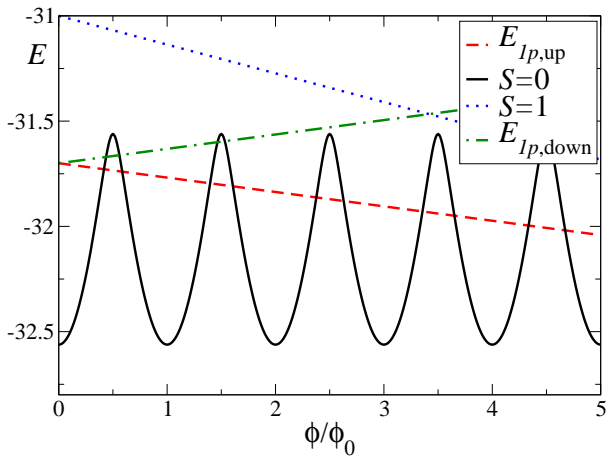}
\caption{(Color online) Energy levels of the system as a function of the
applied magnetic flux.}
\label{b}
\end{figure}

In Fig. \ref{b} we show the relevant energy levels (for one and two
particles in the system of three QD's) as a function of the applied magnetic
field $B$ for a gate voltage chosen in such a way that there are several
crossings between the ground state for one and two particles. At these
crossings one expects a large conductance \cite{del}, and this is confirmed
by our calculations described below. For two particles, the ground state is
a singlet except for magnetic flux $\phi $ larger than 10 times the flux
quantum $\phi _{0}=h/e$, for which the ground state becomes the triplet with
projection 1 \cite{note}.

\begin{figure}[h]
\includegraphics[width=8cm]{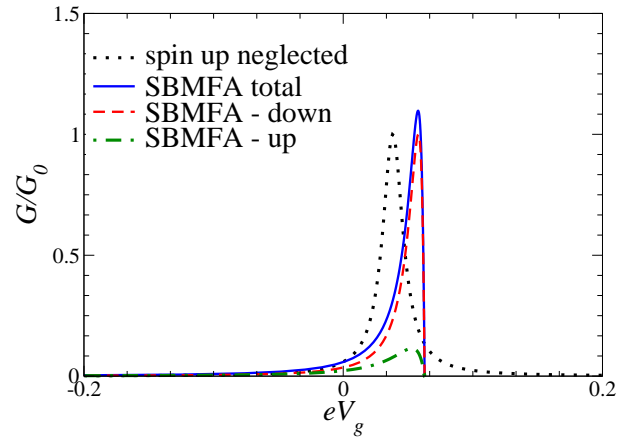}
\caption{(Color online) Conductance for each spin as a function of gate
voltage of the effective Anderson model for $t_{LD}=4.0$ and $\protect\phi /%
\protect\phi _{0}= 0.43$.}
\label{bos}
\end{figure}

The one-particle states with lowest energy are Kramers degenerate for $B=0$
and are split by the Zeeman term for non-zero applied magnetic field.
However, for small $B$, in principle both states should be retained, and the
procedure used to derive Eq. (\ref{cond}) is in principle invalid. To test
this, we have calculated the conductance of the effective impurity Anderson
model (which hybridizes a doublet and a singlet via electronic transitions
to or from the leads) in the SBMFA \cite{ihm,lady}, mentioned briefly in the
previous section. In Fig. \ref{bos} we have selected the magnetic flux that
corresponds to the first crossing between the lowest energy levels for one
and two particles, and for this magnetic field, we represent the conductance
as a function of gate voltage. The contribution for spin down dominates the
conductance. For both spins there is a peak in the conductance near the gate
voltage for which an electron can be taken from the Fermi energy and added
to the system with one particle, forming the singlet, without cost of
energy. The asymmetry of the peaks, with an abrupt fall to zero at the
right, is an artifact of the SBMFA. The dotted line corresponds to Eq. (\ref%
{cond}) in which the contribution to the conductance of the one-particle
level with spin up is neglected. We can see that the contribution of the
level with spin up is small even in this case.

This result confirms that when the Zeeman splitting $g \mu_B B$ is much
larger than $\Delta =2$ $\tilde{t}_{LD}^{2}/t_{L}$, the approximation of
taking only the lowest one-particle level is valid. For the case of Fig. \ref%
{bos} one has $g \mu_B B=0.117$ and $\Delta=0.098$. For all other crossings
studied here, either the splitting of the one particle states is larger, or $%
t_{LD}$ is smaller than 4 (reducing $\Delta$). Therefore, we use this
approximation, leading to Eq. (\ref{cond}) in the rest of this work. This is
not only simpler, but also more accurate than the SBMFA when $g \mu_B B \gg
\Delta$, in particular due to the above mentioned shortcoming of the SBMFA.

\begin{figure}[h]
\includegraphics[width=8cm]{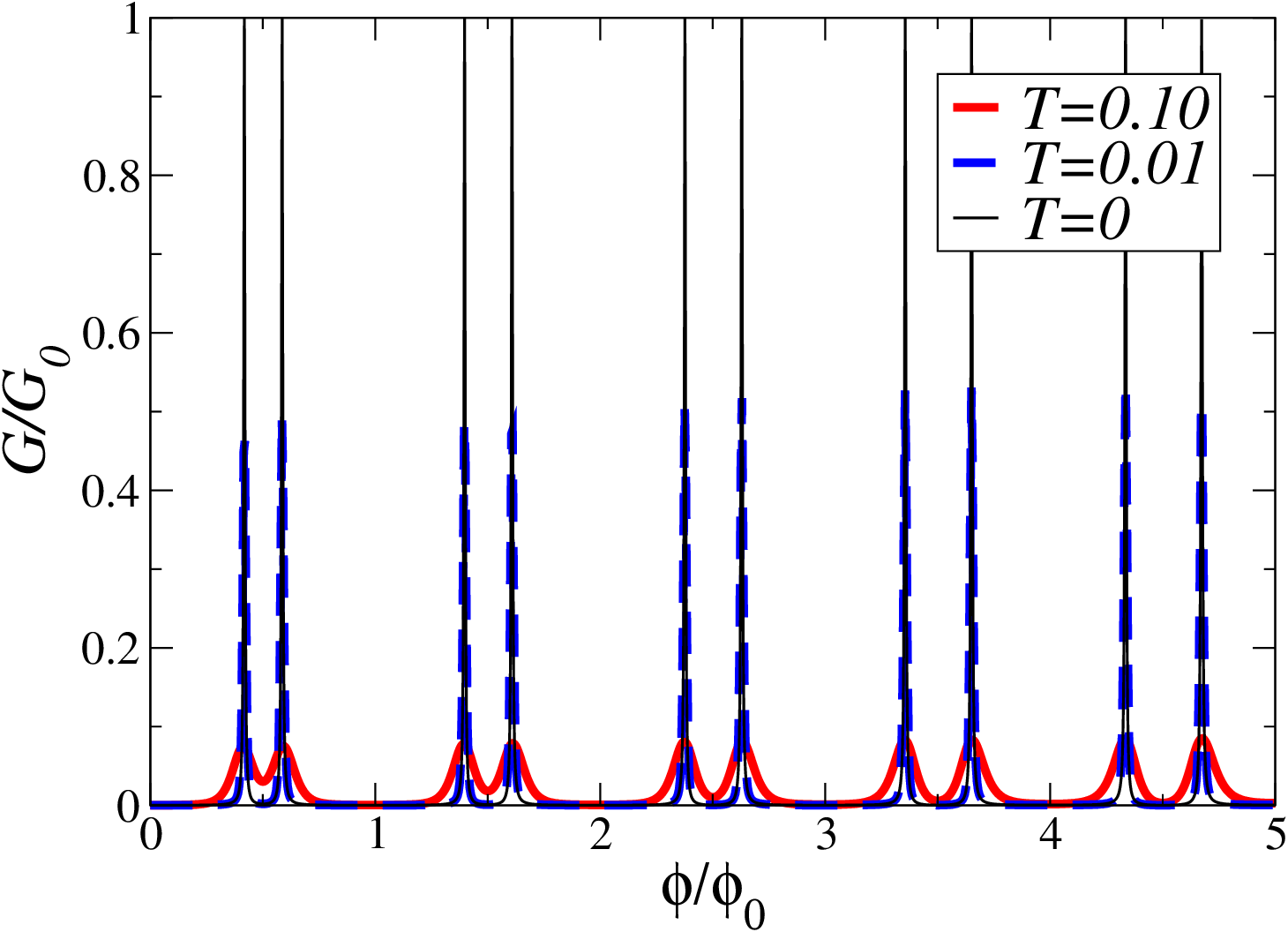}
\caption{(Color online) Conductance as a function of the applied magnetic
flux for different temperatures and for $t_{LD}=2$.}
\label{e}
\end{figure}

Using Eq. (\ref{cond}), taking $t_{LD}=2$ and the rest of the parameters as
in Fig. \ref{b}, we obtain the conductance as a function of flux shown in
Fig. \ref{e}. As expected, there are peaks at the values of the flux which
correspond to the crossing points in Fig. \ref{b}. Our results are in
qualitative agreement with those of Delgado \textit{et al.}

In addition, our approach allows us to show that at zero temperature the
height of the peaks reaches the maximum possible for given spin $%
G_{0}=e^{2}/h$. This ideal conductance corresponds to the symmetric case we
assumed in which both leads are coupled to the dots in the same way. If not,
the maximum conductance is reduced by a well known factor $A$ which depends
on the difference between left and right effective couplings $\Delta$ at the
Fermi energy \cite{rin,pro}.

In Fig. \ref{f} we show how the results are modified when $t_{LD}$ is
increased by a factor 2, leading to a four times larger resonant level width 
$\Delta =2$ $\tilde{t}_{LD}^{2}/t_{L}$. As expected, the peaks are broadened
and are more robust under the effect of temperature. These features might
allow experimentalists an easier tuning of a spin filter device. We note
that for this value $t_{LD}=4$ or higher ones, the approach of Ref. \cite%
{del} predicts a conductance per spin \emph{higher} than $G_0$. We must warn
the reader that the conductance for $T=0.1$ (near 25 Kelvin) and flux near
half a flux quantum are actually an underestimation, because the energy of
the thermal excitation $kT$ is of the order of the Zeeman splitting $g \mu_B
B$ and therefore both spin states contribute to the conductance.

\vspace{1.0cm} 
\begin{figure}[h]
\includegraphics[width=8cm]{f.eps}
\caption{(Color online) Same as Fig. \protect\ref{e} for $t_{LD}=4$.}
\label{f}
\end{figure}

\subsection{System with spin-orbit coupling}

\begin{figure}[h]
\includegraphics[width=8cm]{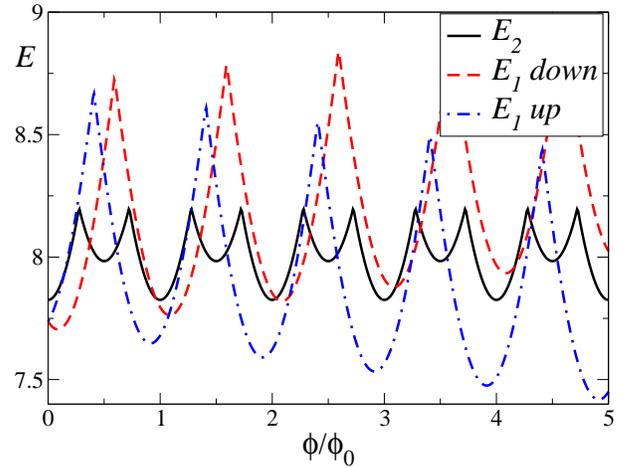}
\caption{(Color online) Relevant energy levels as a function of the applied
magnetic flux.}
\label{a}
\end{figure}

Here we report the results with $\alpha \neq 0$, and $E_{1}=E_{2}=E_{3}$. In
Fig. \ref{a}, we show the energy of the ground state for two electrons, and
the energy of the lowest-lying one-particle states, for a spin-orbit
coupling such that $R=2$. The total spin ceases to be a good quantum number,
but the ground state for two particles is adiabatically connected to a state
that is a singlet for $R=0$, and is well separated from the states of higher
energy. In contrast to the previous case, the one-particle levels now have
an oscillating structure as a function of the applied magnetic field, in
addition to the Zeeman splitting. This renders the maps of the crossing
points more involved, with quasi-periodic changes in the spin of the
one-particle ground state as the flux increases.

\begin{figure}[h]
\includegraphics[width=8cm]{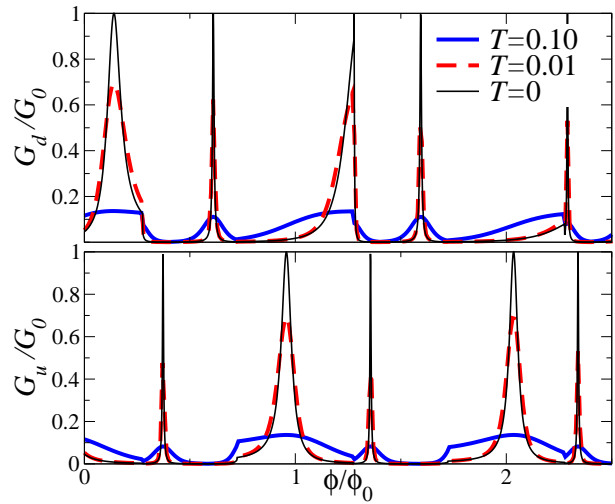}
\caption{(Color online) Conductance as a function of the applied magnetic
flux for different temperatures, $t_{LD}=2$ and $R=2$. Top (bottom)
corresponds to spin down (up).}
\label{d}
\end{figure}

In any case, except at some particular points \cite{note2}, there is a spin
selective conductance and as shown in Fig. \ref{d}, the peaks in which the
conductance is only up or down (in the appropriate quantization axis)
alternate as a function of the applied magnetic field. The broader peaks
correspond to crossings of the energy levels in which the slope of the one-
and two-particle levels as a function of the flux is more similar. For
example, in Fig. \ref{a}, it is clear that the crossing between the energy
for spin up and the two-particle state at a magnetic flux near $0.6 \phi _{0}
$ is more abrupt than the corresponding crossing at $\phi \sim 0.2 \phi _{0}$%
, and as a consequence, the peak in the conductance at the former crossing
is sharper (see Fig. \ref{d}).

\begin{figure}[h]
\includegraphics[width=8cm]{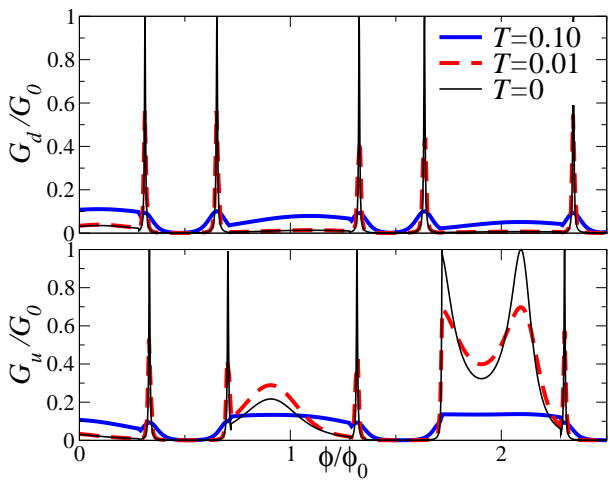}
\caption{(Color online) Same as Fig. \protect\ref{d} for $R=1$}
\label{i}
\end{figure}

In Fig. \ref{i} we show the changes in the conductance for a smaller value
of the spin-orbit coupling $R=1$. While the same qualitative features are
retained, it becomes more difficult to separate the values of the flux for
which there is a peak in the conductance for spin up or down, for small
applied magnetic field. This is due to the fact that the one-particle
energies for both spin directions are more similar for small applied
magnetic field. For larger fields, both energies become well separated by
the Zeeman term, as in the previous case.

\section{Summary and discussion}

We have investigated a system of three QD's taking essentially parameters
estimated previously \cite{del}, for which the system acts as a magnetic
field tunable spin valve. 
A significant spin-valve effect should be observable when either one spin is
localized at the QD not connected to the leads because of its lower on-site
energy combined with strong on-site Coulomb repulsion \cite{del}.
Experimentally, it has been show that it is possible to control independently
the on-site energies by different gate voltages (see for example Ref. cite{rogge}).
Another possibility to obtain a spin filtering effect is to thread the system
by a magnetic flux under
the effect of spin-orbit coupling \cite{abc,abc2}. In this case, as
the magnetic flux is changed, spikes in the conductance for opposite spin
orientations alternate, a fact that may be of interest for applications.
For simplicity, in the calculation which includes spin-orbit coupling, we
have assumed the same on-site energy at the three dots (point group $C_{3}$). 
However, our conclusions are not altered if the difference between
on-site energies is of the order of the interdot hopping.  

While spin filtering effects in arrays of quantum dots has been studied
previously \cite{abc,abc2,oje,del,chi,chi2,torio,lady,sun,heary,vernek}, the
fact that the effective resonant level width $\Delta $ is in general (for
the given parameters) much smaller that the separation between the energy
levels in the system, allows us to map the problem onto a much simpler,
noninteracting one, for which the conductance can be calculated without
further approximations. In particular, in spite of its simplicity, our
results provide a significant quantitative improvement on those of Delgado 
\textit{et al.} \cite{del,del2}, and agree with the fact that the maximum
conductance per spin is the ideal one for the system $G_{0}$ independently
of the lead-dot hopping $t_{LD}$.

In fact, from our study we conclude that the main ingredients to have a good
spin-filtering effect is to break the Kramers degeneracy (either by an
applied magnetic field on localized spins or by a spin-orbit coupling), in
such a way that the separation between the low-lying levels for a
configuration with $n$ electrons is larger than $\Delta$. This allows in
general, the mapping to a non-interacting model. A peak in the conductance
is obtained when the gate voltage is such that electrons can be transfered
between both configurations without energy cost.

The most important correlations in the problem are contained in the
description of the isolated system of three dots. The hybridization of the
system with the leads in general introduces new effects of correlations,
related with the Anderson and Kondo physics \cite{ihm,abc,abc2}. However, we
find that these effects are minor for the parameters used, particularly due
to the breaking of the Kramers degeneracy.

Another advantage of the effective non-interacting model is that the
conductance can be calculated easily in a non-equilibrium situation, as long
as the applied bias voltage is small compared to the separation between
low-lying energy levels in both configurations. Instead, to include
non-equilibrium effects in strongly correlated models is very difficult at
present \cite{none}.

\section*{Acknowledgments}

One of us (AAA) is partially supported by CONICET. This work was done in the
framework of projects PIP 11220080101821 of CONICET, and PICT 2006/483 and
PICT R1776 of the ANPCyT.


\begin{thebibliography}{99}
\bibitem{pett} J. R. Petta, A. C. Johnson, J. M. Taylor, E. A. Laird, A.
Yacoby, M. D. Lukin, C. M. Marcus, M. P. Hanson, and A. C. Gossard, Science 
\textbf{309}, 2180 (2005).

\bibitem{grei} A. Greilich, D. R. Yakovlev, A. Shabaev, Al. L. Efros, I. A.
Yugova, R. Oulton, V. Stavarache, D. Reuter, A. Wieck, and M. Bayer, Science 
\textbf{313}, 341 (2006).

\bibitem{ata} M. Atat\"{u}re, J. Dreiser, A. Badolato, A. H\"{o}gele, K.
Karrai, and A. Imamoglu, Science \textbf{312}, 551 (2006).

\bibitem{grei2} A. Greilich, R. Oulton, E. A. Zhukov, I. A. Yugova, 
D. R. Yakovlev, M. Bayer, A. Shabaev, Al. L. Efros, I. A. Merkulov, V. Stavarache,
D. Reuter, and A. Wieck, Phys. Rev. Lett. \textbf{96}, 227401 (2006).

\bibitem{bere} J. Berezovsky, M. H. Mikkelsen, N. G. Stoltz, L. A. Caldren,
and D. D. Awshalom, Science \textbf{320}, 349 (2008).

\bibitem{mac} S. Mackowski, T. Gurung, H. E. Jackson, L. M. Smith, G.
Karczewski and J. Kossut, Appl. Phys. Lett. \textbf{87}, 072502 (2005).

\bibitem{kork} M. Korkusinski and P. Hawrylak, Phys. Rev. Lett. \textbf{101}%
, 027205 (2008).

\bibitem{rei} D. E. Reiter, T. Kuhn, and V. M. Axt, Phys. Rev. Lett. \textbf{%
102}, 177403 (2009).

\bibitem{gold1} D. Goldhaber-Gordon, H. Shtrikman, D. Mahalu, D.
Abusch-Magder, U. Meirav, and M. A. Kastner, Nature \textbf{391}, 156 (1998).

\bibitem{cro} S. M. Cronenwet, T. H. Oosterkamp, and L. P. Kouwenhoven,
Science \textbf{281}, 540 (1998).

\bibitem{gold2} D. Goldhaber-Gordon, J. G\"{o}res, M. A. Kastner, H.
Shtrikman, D. Mahalu, and U. Meirav, Phys. Rev. Lett. \textbf{81}, 5225
(1998).

\bibitem{wiel} W. G. van der Wiel, S. de Franceschi, T. Fujisawa, J. M.
Elzerman, S. Tarucha, and L. P. Kowenhoven, Science \textbf{289}, 2105
(2000).

\bibitem{man} H. C. Manoharan, C. P. Lutz, and D. M. Eigler, Nature (London) 
{\ \textbf{403}, 512 (2000); references therein.}

\bibitem{glaz} L.I. Glazman and M.E. Raikh, JETP Lett. \textbf{47}, 452
(1988).

\bibitem{ng} T. K. Ng and P. A. Lee, Phys. Rev. Lett. \textbf{61}, 1768
(1988).

\bibitem{chz} T. A. Costi, A. C. Hewson, and V. Zlati\'{c}, J. Phys.
Condens. Matter \textbf{6}, 2519 (1994).

\bibitem{rev} A. A. Aligia and A. M. Lobos, J. Phys.: Condens. Matter 
\textbf{17}, S1095 (2005); references therein.

\bibitem{grobis} M. Grobis, I. G. Rau, R. M. Potok, H. Shtrikman, and D.
Goldhaber-Gordon, Phys. Rev. Lett. \textbf{100}, 246601 (2008)..

\bibitem{rin} J. Rinc\'{o}n, A. A. Aligia, K. Hallberg, Phys. Rev. B \textbf{%
79}, 121301(R) (2009); \textbf{80}, 079902(E) (2009); \textbf{81}, 039901(E)
(2010).

\bibitem{scott} G. D. Scott, Z. K. Keane, J. W. Ciszek, J. M. Tour, and D.
Natelson, Phys. Rev. B \textbf{79}, 165413 (2009).

\bibitem{sela} E. Sela and J. Malecki, Phys. Rev. B \textbf{80}, 233103
(2009).

\bibitem{roch} N. Roch, S. Florens, V. Bouchiat, W. Wernsdorfer, and F.
Balestro, Nature (London) \textbf{453}, 633 (2008).

\bibitem{rou1} P. Roura Bas and A. A. Aligia, Phys. Rev. B \textbf{80},
035308 (2009).

\bibitem{logan} D. E. Logan, C. J. Wright, and M. R. Galpin, Phys. Rev. B 
\textbf{80}, 125117 (2009)

\bibitem{roch2} N. Roch, S. Florens, T. A. Costi, W. Wernsdorfer, and F.
Balestro, Phys. Rev. Lett. \textbf{103}, 197202 (2009).

\bibitem{rou2} P. Roura Bas and A. A. Aligia, J. Phys. Cond. Matt. \textbf{22%
}, 025602 (2010).

\bibitem{mitc} A. K. Mitchell, T. F. Jarrold, and D. E. Logan, Phys. Rev. B 
\textbf{79}, 085124 (2009).

\bibitem{dal} P. A. Dalgarno, M. Ediger, B. D. Gerardot, J. M. Smith, S.
Seidl, M. Kroner, K. Karrai, P. M. Petroff, A. O. Govorov, and R. J.
Warburton, Phys. Rev. Lett. \textbf{100}, 176801 (2008)

\bibitem{misael} L. M. Le\'{o}n Hilario and A. A. Aligia, Phys. Rev. Lett. 
\textbf{103}, 156802 (2009).

\bibitem{jamn} T. Jamneala, V. Madhavan, and M. F. Crommie, Phys. Rev. Lett. 
\textbf{87}, 256804 (2001).

\bibitem{trimer} A. A. Aligia, Phys. Rev. Lett. \textbf{96}, 096804 (2006);
references therein.

\bibitem{craig} N. J. Craig, J. M. Taylor, E. A. Lester, C. M. Marcus, M. P.
Hanson, and A. C. Gossard, Science \textbf{293}, 2221 (2001).

\bibitem{hol} A.W. Holleitner, C. R. Decker, H. Qin, K. Eberl, and R. H.
Blick, Phys. Rev. Lett. \textbf{87}, 256802 (2001).

\bibitem{jeong} H. Jeong, A. M. Chang, and M. R. Meloch, Science \textbf{304}%
, 565 (2004).

\bibitem{chen} J. C. Chen, A. M. Chang, and M. R. Melloch, Phys. Rev. Lett. 
\textbf{92}, 176801 (2004).

\bibitem{vidan} A. Vidan, R. M. Westervelt, M. Stopa, M. Hanson, and A. C.
Gossard, Appl. Phys. Lett. \textbf{85}, 3602 (2004).

\bibitem{petta} J. R. Petta, A. C. Johnson, J. M. Taylor, E. A. Laird, A.
Yacoby, M. D. Lukin, C. M. Marcus, M. P. Hanson, and A. C. Gossard, Science 
\textbf{309}, 2180 (2005) 

\bibitem{waugh} F. R. Waugh, M. J. Berry, D. J. Mar, R. M. Westervelt, K. L.
Campman, and A. C. Gossard, Phys. Rev. Lett. \textbf{75}, 705 (1995).

\bibitem{gaud} L. Gaudreau, S. A. Studenikin, A. S. Sachrajda, P. Zawadzki,
A. Kam, J. Lapointe, M. Korkusinski, and P. Hawrylak, Phys. Rev. Lett. 
\textbf{97}, 036807 (2006).

\bibitem{gaudpss} L. Gaudreau, A. S. Sachrajda, S. Studenikin, P. Zawadzki,
J. Lapointe, and A. Kam, Phys. Status Solidi C \textbf{3}, 3757 (2006).

\bibitem{gaudprb} L. Gaudreau, A. S. Sachrajda, S. Studenikin, A. Kam, F.
Delgado, Y. P. Shim, M. Korkusinski, and P. Hawrylak, Phys. Rev. B \textbf{80%
}, 075415 (2009). 

\bibitem{grove} K. Grove-Rasmussen, H. I. J\o rgensen, T. Hayashi, P. E.
Lindelof, and T. Fujisawa, Nano Lett. \textbf{8}, 1055 (2008).

\bibitem{schro} D. Schr\"{o}er, A. D. Greentree, L. Gaudreau, K. Eberl, L.
C. L. Hollenberg, J. P. Kotthaus, and S. Ludwig, Phys. Rev. B \textbf{76},
075306 (2007).

\bibitem{rogge} M. C. Rogge and R. J. Haug, Phys. Rev. B \textbf{77}, 193306
(2008). 

\bibitem{kouw} L. P. Kouwenhoven, F. W. J. Hekking, B. J. van Wees, C. J. P.
M. Harmans, C. E. Timmering, and C. T. Foxon, Phys. Rev. Lett. \textbf{65},
361 (1990).

\bibitem{dias} L. G. G. V. Dias da Silva, N. P. Sandler, K. Ingersent, and
S. E. Ulloa, Phys. Rev. Lett. \textbf{97}, 096603 (2006); \textit{ibid} 
\textbf{99}, 209702 (2007); L. Vaugier, A.A. Aligia and A.M. Lobos, \textit{%
ibid} \textbf{99}, 209701 (2007); L. Vaugier, A.A. Aligia and A.M. Lobos,
Phys. Rev. B \textbf{76}, 165112 (2007).

\bibitem{oje} J. H. Ojeda, M. Pacheco, and P. A. Orellana, Nanotechnology 
\textbf{20}, 434013 (2009).

\bibitem{oreg} Y. Oreg and D. Goldhaber-Gordon, Phys. Rev. Lett. \textbf{90}%
, 136602 (2003).

\bibitem{zit} R. \v{Z}itko and J. Bon\v{c}a, Phys. Rev. B \textbf{74},
224411 (2006).

\bibitem{mart} G. B. Martins, C. A. B\"{u}sser, K. A. Al-Hassanieh, A.
Moreo, and E. Dagotto, Phys. Rev. Lett. \textbf{94}, 026804 (2005).

\bibitem{ihm} A. A. Aligia, K. Hallberg, B. Normand, and A. P. Kampf, Phys.
Rev. Lett. \textbf{93}, 076801 (2004).

\bibitem{corn} P. S. Cornaglia and D. R. Grempel, Phys. Rev. B \textbf{71},
075305 (2005).

\bibitem{zit2} R. \v{Z}itko and J. Bon\v{c}a, Phys. Rev. B \textbf{73},
035332 (2006).

\bibitem{ogur} A. Oguri , Y. Nisikawa, and A. C. Hewson, J. Phys. Soc. Jpn. 
\textbf{74}, 2554 (2005).

\bibitem{nisi} Y. Nisikawa and A. Oguri, Phys. Rev. B \textbf{73}, 125108
(2006).

\bibitem{3do} A. M. Lobos and A. A. Aligia, Phys. Rev. B \textbf{74}, 165417
(2006).

\bibitem{kuzme} T. Kuzmenko, K. Kikoin, and Y. Avishai, Phys. Rev. Lett. 
\textbf{96}, 046601 (2006).

\bibitem{dinu} I. V. Dinu, M. \c{T}olea, and A. Aldea, Phys. Rev. B \textbf{%
76}, 113302 (2007).

\bibitem{ldeca} E. V. Anda, G. Chiappe, C. A. B\"{u}sser, M. A. Davidovich,
G. B. Martins, F. Heidrich-Meisner, and E. Dagotto, Phys. Rev. B \textbf{78}%
, 085308 (2008).

\bibitem{abc} A. M. Lobos and A. A. Aligia, Phys. Rev. Lett. \textbf{100},
016803 (2008).

\bibitem{abc2} A. M. Lobos and A. A. Aligia, Physica B \textbf{404}, 3306
(2009).

\bibitem{del} F. Delgado, Y.-P. Shim, M. Korkusinski, L. Gaudreau, S. A.
Studenikin, A. S. Sachrajda, and P. Hawrylak, Phys. Rev. Lett. \textbf{101},
226810 (2008).

\bibitem{del2} Y.-P. Shim, F. Delgado, and P. Hawrylak,  Phys. Rev. B 
\textbf{80}, 115305 (2009).

\bibitem{del3} F. Delgado and P. Hawrylak, J. Phys.: Condens. Matter \textbf{%
20}, 315207 (2008).

\bibitem{mitc2} A. K. Mitchell and D. E. Logan, Phys. Rev. B \textbf{81},
075126 (2010). 

\bibitem{sq} A. Ram\v{s}ak, J. Mravlje, R. \v{Z}itko and J. Bon\v{c}a, Phys.
Rev. B \textbf{74}, 241305(R) (2006).

\bibitem{zaf} M. Zaffalon, A. Bid, M. Heiblum, D. Mahalu, and V. Umansky,
Phys. Rev. Lett. \textbf{100}, 226601 (2008); references therein.

\bibitem{ji} Y. Ji, M. Heiblum, D. Sprinzak, D. Mahalu, and H. Shtrikman,
Science \textbf{290}, 779 (2000).

\bibitem{bayer} M. Bayer, M. Korkusinski, P. Hawrylak, T. Gutbrod, M.
Michel, and A. Forchel, Phys. Rev. Lett. \textbf{90}, 186801 (2003). 

\bibitem{teodo} M. D. Teodoro, V. L. Campo, Jr., V. Lopez-Richard, E.
Marega, Jr., G. E. Marques, Y. Galv\~{a}o Gobato, F. Iikawa, M. J. S. P.
Brasil, Z. Y. AbuWaar, V. G. Dorogan, Yu. I. Mazur, M. Benamara, and G. J.
Salamo,  Phys. Rev. Lett. \textbf{104}, 086401 (2010).

\bibitem{kon} M. K{\"{o}}nig, A. Tschetschetkin, E. M. Hankiewicz, J.
Sinova, V. Hock, V. Daumer, M. Sch\"{a}fer, C. R. Becker, H. Buhmann, and L.
W. Molenkamp, Phys. Rev. Lett. \textbf{96}, 076804 (2006).

\bibitem{ber} T. Bergsten, T. Kobayashi, Y. Sekine, and J. Nitta, Phys. Rev.
Lett. \textbf{97}, 196803 (2006).

\bibitem{ac} Y. Aharonov and A. Casher, Phys. Rev. Lett. \textbf{53}, 319
(1984).

\bibitem{shen} S.-Q. Shen, Z.-J. Li, and Z. Ma, Appl. Phys. Lett. \textbf{84}%
, 996 {2004). }

\bibitem{mol} B. Moln\'{a}r, F. M. Peeters, and P. Vasilopoulos, Phys. Rev.
B \textbf{69}, 155335 {2004).}

\bibitem{chi} F. Chi, J. L. Liu, and L. L. Sun, J. Appl. Phys. \textbf{101},
093704 (2007)

\bibitem{chi2} F. Chi and S. S. Li, J. Appl. Phys. \textbf{100}, 113703
(2006).

\bibitem{torio} M. E. Torio, K. Hallberg, S. Flach, A. E. Miroshnichenko,
and M. Titov, Eur. Phys. J. B 37, 399 (2004).

\bibitem{lady} A. A. Aligia and L. A. Salguero, Phys. Rev. B \textbf{70},
075307 (2004); \textbf{71}, 169903(E) (2005).

\bibitem{sun} Q.-F. Sun, J. Wang, and H. Guo, Phys. Rev. B \textbf{71},
165310 (2005).

\bibitem{heary} R. J. Heary, J. E. Han, and L. Zhu, Phys. Rev. B \textbf{77}%
, 115132 (2008).

\bibitem{vernek} E. Vernek, N. Sandler, and S. E. Ulloa, Phys. Rev. B 
\textbf{80}, 041302(R) (2009).

\bibitem{busl} Maria Busl, Rafael S\~{A}\textexclamdown nchez, and Gloria
Platero, Phys. Rev. B \textbf{81}, 121306(R) (2010).

\bibitem{schm} D. Schmeltzer, A. R. Bishop, A. Saxena, and D. L. Smith,
Phys. Rev. Lett. \textbf{90}, 116802 (2003).

\bibitem{rist} Z. Ristivojevic, G. I. Japaridze, and T. Nattermann, Phys.
Rev. Lett. \textbf{104}, 076401 (2010).

\bibitem{meir} Y. Meir and N. S. Wingreen, Phys. Rev. Lett. \textbf{68},
2512 (1992).

\bibitem{inter} J. Rinc\'{o}n, A. A. Aligia, and K. Hallberg, Phys. Rev. B 
\textbf{79}, 035112 (2009).

\bibitem{hews} A. C. Hewson, in \textit{The Kondo Problem to Heavy Fermions }%
(Cambridge, University Press, 1993).

\bibitem{none} A. A. Aligia, Phys. Rev. B \textbf{74}, 155125 (2006);
references therein.

\bibitem{note} For high enough magnetic field $B$, the ground state for two
particles $|2g\rangle$ becomes a triplet with maximum spin projection if $%
\alpha=0$. The derivation of the appropriate effective Hamiltonian for this
case, is similar to the one given in the text.

\bibitem{pro} A.A. Aligia and C.R. Proetto, Phys. Rev. B \textbf{65}, 165305
(2002).

\bibitem{gim} I. Puerto Gimenez, M. Korkusinski, and P. Hawrylak, Phys. Rev.
B \textbf{76}, 075336 (2007)

\bibitem{usaj} G. Usaj and C. A. Balseiro, Phys. Rev. B \textbf{70},
041301(R) (2004).

\bibitem{miller} J. B. Miller, D. M. Zumb\"uhl, C. M. Marcus, Y. B.
Lyanda-Geller, D. Goldhaber-Gordon, K. Campman, and A. C. Gossard, Phys.
Rev. Lett. \textbf{90} 076807 (2003).

\bibitem{nitta} J. Nitta, T. Akazaki, H. Takayanagi, and T. Enoki, Phys.
Rev. Lett. \textbf{78}, 1335 (1997).

\bibitem{grun} D. Grundler, Phys. Rev. Lett. \textbf{84}, 6074 (2000).

\bibitem{note2} For those special points in which the three energy levels
shown in Fig. \ref{a} are nearly degenerate, the effective Hamiltonian (\ref%
{heff}) becomes invalid and a more elaborate formalism shold be used to
calculate the conductance.\cite{abc}
\end{thebibliography}
\end{document}